\documentclass[manuscript]{emulateapj}

\newcommand{\secp}{\mbox{\rlap{.}$''$}} 
\newcommand{\rasecp}{\mbox{\rlap{.}$^{\rm s}$}}

\newcommand{\jms}{J. Mol. Spectr.}

\newcommand{\pccp}{Phys. Chem. Chem. Phys.}

\newcommand{\nature}{Nature}
\newcommand{\jcppcb}{J. Chim. Phys. Phys.-Chim. Biol.}
\newcommand{\jpcrd}{J. Phys. Chem. Ref. Data}

\shorttitle{The distribution of CH$_3$CN in IRC\,+10216}
\shortauthors{Ag\'undez et al.}

\begin{document} 

\title{The peculiar distribution of CH$_3$CN in IRC\,+10216 seen by ALMA\thanks{Based on observations carried out with ALMA and the IRAM 30m Telescope. ALMA is a partnership of ESO (representing its member states), NSF (USA) and NINS (Japan), together with NRC (Canada) and NSC and ASIAA (Taiwan), in cooperation with the Republic of Chile. The Joint ALMA Observatory is operated by ESO, AUI/NRAO and NAOJ. IRAM is supported by INSU/CNRS (France), MPG (Germany) and IGN (Spain). This paper makes use of the following ALMA data: ADS/JAO.ALMA\#2011.0.00229.S.}}

\author{M.~Ag\'undez$^1$, J.~Cernicharo$^1$, G.~Quintana-Lacaci$^1$, L.~Velilla Prieto$^1$, A.~Castro-Carrizo$^2$, N.~Marcelino$^3$, and M.~Gu\'elin$^2$} 

\affil{
$^1$ Instituto de Ciencia de Materiales de Madrid, CSIC, C/ Sor Juana In\'es de la Cruz 3, 28049 Cantoblanco, Spain \\
$^2$ Institut de Radioastronomie Millim\'etrique, 300 rue de la Piscine, 38406 St. Martin d'H\'eres, France \\
$^3$ INAF, Istituto di Radioastronomia, via Gobetti 101, 40129 Bologna, Italy
}

\begin{abstract}

IRC\,+10216 is a circumstellar envelope around a carbon-rich evolved star which contains a large variety of molecules. According to interferometric observations, molecules are distributed either concentrated around the central star or as a hollow shell with a radius of $\sim$15$''$. We present ALMA Cycle\,0 band\,6 observations of the $J=14-13$ rotational transition of CH$_3$CN in IRC\,+10216, obtained with an angular resolution of $0\secp76\times0\secp61$. The bulk of the emission is distributed as a hollow shell located at just $\sim$2$''$ from the star, with a void of emission in the central region up to a radius of $\sim$1$''$. This spatial distribution is markedly different from those found to date in this source for other molecules. Our analysis indicate that methyl cyanide is not formed neither in the stellar photosphere nor far in the outer envelope, but at radial distances as short as 1-2$''$, reaching a maximum abundance of $\sim0.02$~molecules~cm$^{-3}$ at 2$''$ from the star. Standard chemical models of IRC\,+10216 predict that the bulk of CH$_3$CN molecules should be present at a radius of $\sim15''$, where other species such as polyyne radicals and cyanopolyynes are observed, with an additional inner component within 1$''$ from the star. The non-uniform structure of the circumstellar envelope and grain surface processes are discussed as possible causes of the peculiar distribution of methyl cyanide in IRC\,+10216.

\end{abstract}

\keywords{astrochemistry --- line: identification --- molecular processes --- stars: AGB and post-AGB --- circumstellar matter --- stars: individual (IRC +10216)}

\section{Introduction}

The chemical structure of the well known carbon-rich envelope IRC\,+10216 and, in general, of circumstellar envelopes around asymptotic giant branch (AGB) stars is in general terms well described by an scenario in which stable molecules are formed in the warm and dense surroundings of the star under chemical equilibrium \citep{tsu1973} while radicals and more exotic species are produced in the outer layers due to the photochemistry induced by the penetration of interstellar ultraviolet (UV) photons \citep{gla1996}.

However, in recent years there have been evidences of a growing number of chemical aspects that do not fit into this general scenario. Perhaps, the most clear example is the detection of warm water vapour in IRC\,+10216 and other carbon-rich envelopes \citep{dec2010,neu2011}, as well as the observation of HCN in oxygen-rich envelopes \citep{buj1994,bie2000}, NH$_3$ in O- and C-rich envelopes \citep{kea1993,men2010}, and PH$_3$ in IRC\,+10216 \citep{agu2014}. Observations indicate that these molecules are formed in the inner regions of the envelope with abundances much larger than predicted by chemical equilibrium. Indeed, these inner layers are complex regions where a variety of non-equilibrium processes such as shocks driven by the pulsation of the star \citep{che2012}, processes related to the condensation of dust grains \citep{gai1988}, and even photochemical processes driven by interstellar UV photons able to penetrate through the clumpy envelope \citep{agu2010}, can affect the abundances of some species. Yet, the role of these processes on the chemistry has to be fully understood.

The use of (sub-)mm interferometers able to probe the distribution of different molecules in the inner regions of circumstellar envelopes are a very promising tool to unveil the role of the non-equilibrium processes at work in these inner regions. Some recent works have presented interferometric observations of the inner envelope of IRC\,+10216, revealing the compact emission distribution of some molecules around the star \citep{pat2011,fon2014}. The Atacama Large Millimeter Array (ALMA) has also started to provide observations of the chemical complexity in the inner envelope of IRC\,+10216 with a high angular resolution \citep{cer2013,vel2015}. Here we present ALMA Cycle\,0 band\,6 observations of IRC\,+10216 with sub-arcsec resolution and report a peculiar emission distribution for the $J=14-13$ rotational transition of CH$_3$CN.

\section{Observations} \label{sec:observations}

\begin{figure*}
\centering
\includegraphics[angle=0,width=0.88\columnwidth]{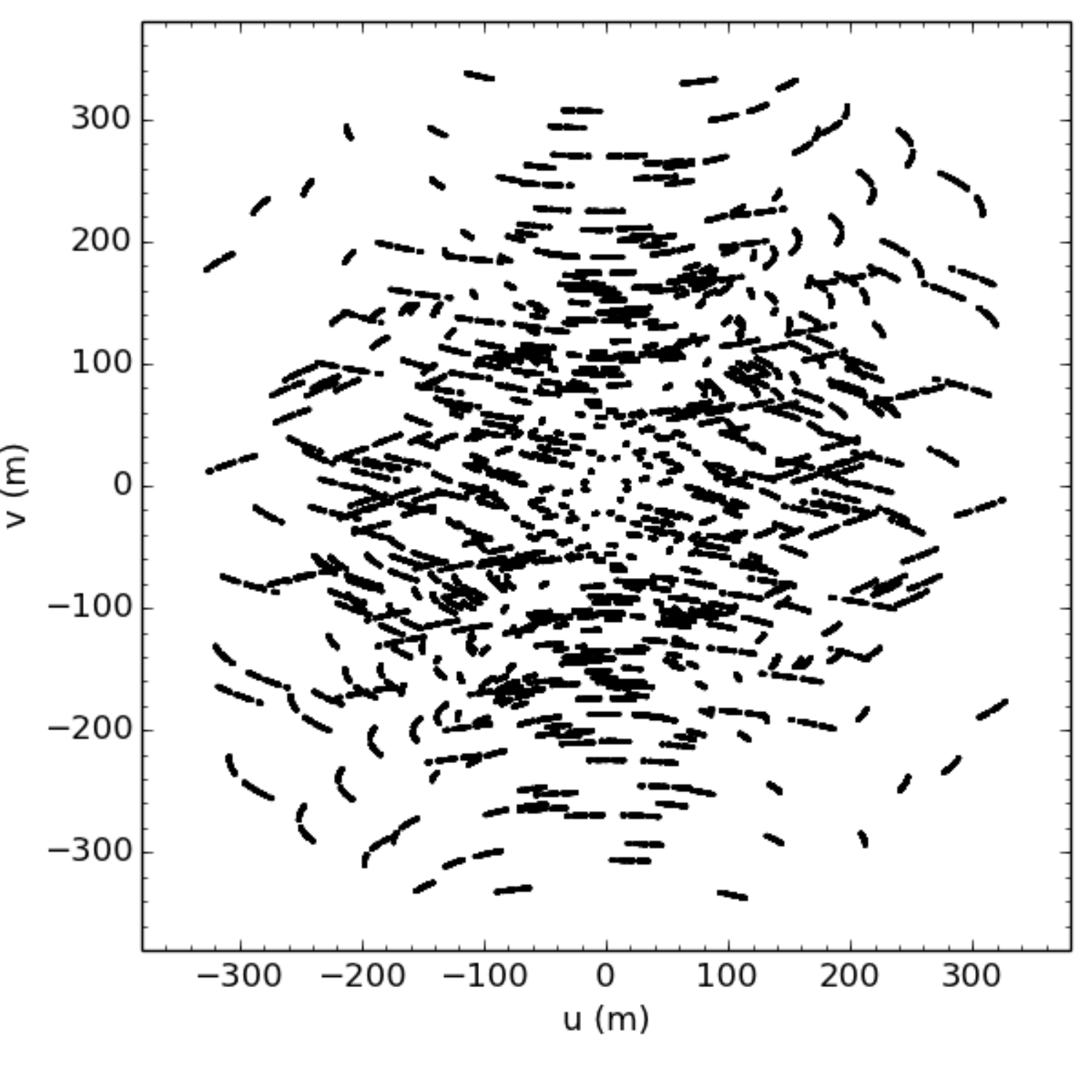} \hspace{1.2cm} \includegraphics[angle=0,width=\columnwidth]{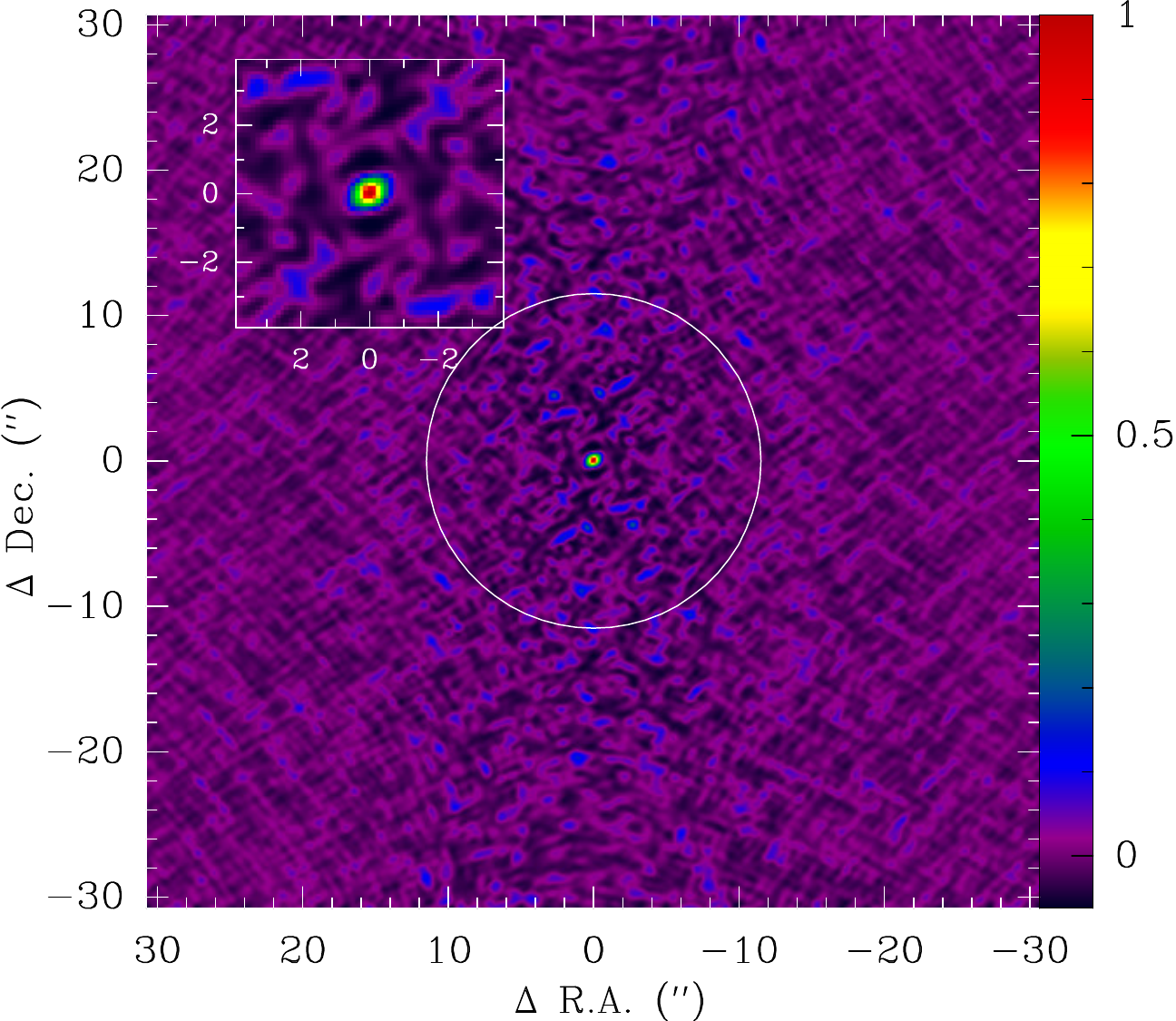}
\caption{ALMA $uv$ plane coverage (left) and associated dirty beam (right) for the spectral setup covering the range 255.3-260.2 GHz. The primary beam of the ALMA 12m antennas at 257.5 GHz ($\sim23''$) is indicated as a white circumference in the right panel.} \label{fig:uv}
\end{figure*}

The ALMA Cycle\,0 observations of IRC\,+10216 were carried out in April 2012 using receiver band\,6. Part of these observations has been previously presented by \citet{cer2013} and by \citet{vel2015}. Here we are concerned with the $J=14-13$ rotational transition of CH$_3$CN, whose strongest $K$ components lie at 257.5 GHz, within the spectral setup covering the 255.3-260.2 GHz frequency range.

An array of 16 antennas with baselines ranging from 15.7~m to 384.1~m was used in three observing tracks of 21 minutes of correlations on source. The field of view (primary beam) of the ALMA 12m antennas has a diameter of $\sim23''$ at 257.5 GHz. The adopted spectral resolution is 0.98 MHz, with a channel spacing of 0.49 MHz. The bright point-like source J0854+201 was observed to calibrate the bandpass, while the amplitude and phase were calibrated by observing J0854+201 and J0909+013 every 10 and 20 minutes, respectively. The uncertainty in the flux calibration is estimated to be 8 \%. Calibration was done with the software CASA\footnote{See \texttt{http://casa.nrao.edu}} and further data processing was done using the package GILDAS\footnote{See \texttt{http://www.iram.fr/IRAMFR/GILDAS}}. A good coverage of the $uv$ plane was obtained, with 156,865 visibilities (see left panel in Fig.~\ref{fig:uv}). The continuum was subtracted by carefully selecting spectral windows free of line emission located in spectral regions near the CH$_3$CN $J=14-13$ lines. The associated dirty beam has minor contributions from sidelobes (less than 10 \% of the main beam, see right panel in Fig.~\ref{fig:uv}). Image deconvolution was carried out using the H\"ogbom clean algorithm with no support, i.e., no a priori structure was assumed for the CH$_3$CN $J=14-13$ brightness distribution. This choice is adequate because the sidelobes of the dirty beam are not important and because CH$_3$CN line emission in IRC\,+10216 does not have a simple structure. The resulting synthesized beam is $0\secp76\times0\secp61$ and the final rms per 0.49 MHz channel is 3.6 mJy beam$^{-1}$.

\begin{figure}[b]
\centering
\includegraphics[angle=0,width=\columnwidth]{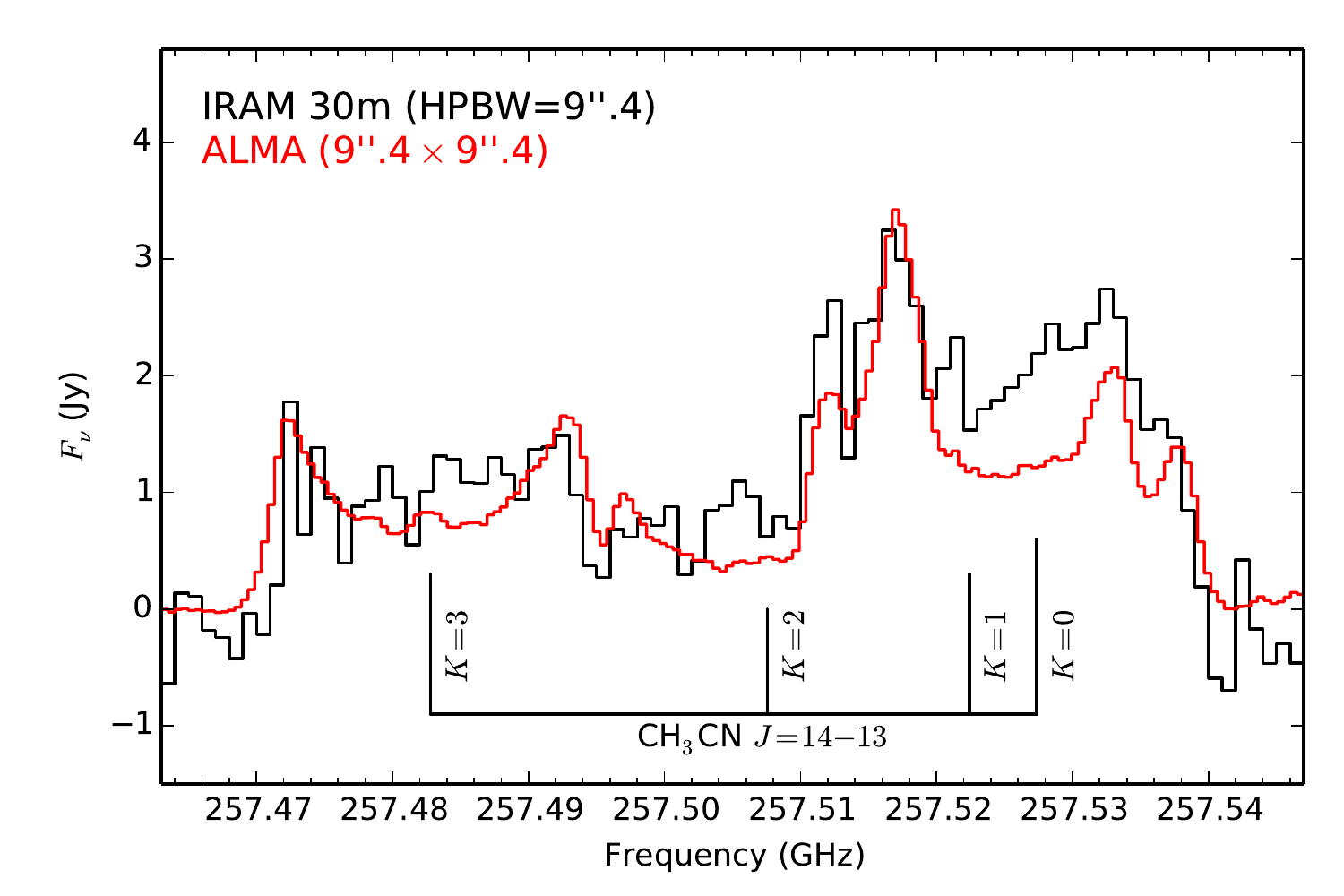}
\caption{Spectrum of IRC\,+10216 in the region of the CH$_3$CN $J=14-13$ transition as observed with the IRAM 30m telescope and with ALMA (integrated over a field of view of similar size to the HPBW of the 30m telescope).} \label{fig:spectrum}
\end{figure}

We carried out observations with the IRAM 30m telescope in June 2014 to estimate the degree of flux that could have been filtered out by ALMA. We used the EMIR E230 receiver connected to a fast Fourier Transform spectrometer providing a spectral resolution of 0.2 MHz (data were later on resampled to the channel spacing of 0.49~MHz of the ALMA data). We performed a scan over a region $44''\times44''$ during 18~h of on source integration time using the `onthefly' observing mode. The half power beam width (HPBW) of the IRAM 30m telescope at 257.5 GHz is 9\secp4. System temperatures were in the range 300-350 K, resulting in a $T_A^*$ rms of $\sim$0.02~K (0.17 Jy) per 0.49~MHz channel in the central region of the map. The spectrum obtained with the IRAM 30m telescope at the position of the star is compared in Fig.~\ref{fig:spectrum} with the ALMA spectrum integrated over a region $9\secp4\times9\secp4$, i.e., similar to the HPBW of the 30m telescope. It is seen that the flux filtered out by ALMA is negligible at the line edges of the different $K$ lines, which correspond to the terminal expansion velocity of the envelope and therefore show a rather compact distribution, although it can be as high as 30-40 \% in some channels located around the center of the $K$ lines, where the emission is more extended.

\section{Results} \label{sec:results}

\begin{figure*}
\centering
\includegraphics[angle=0,width=\textwidth]{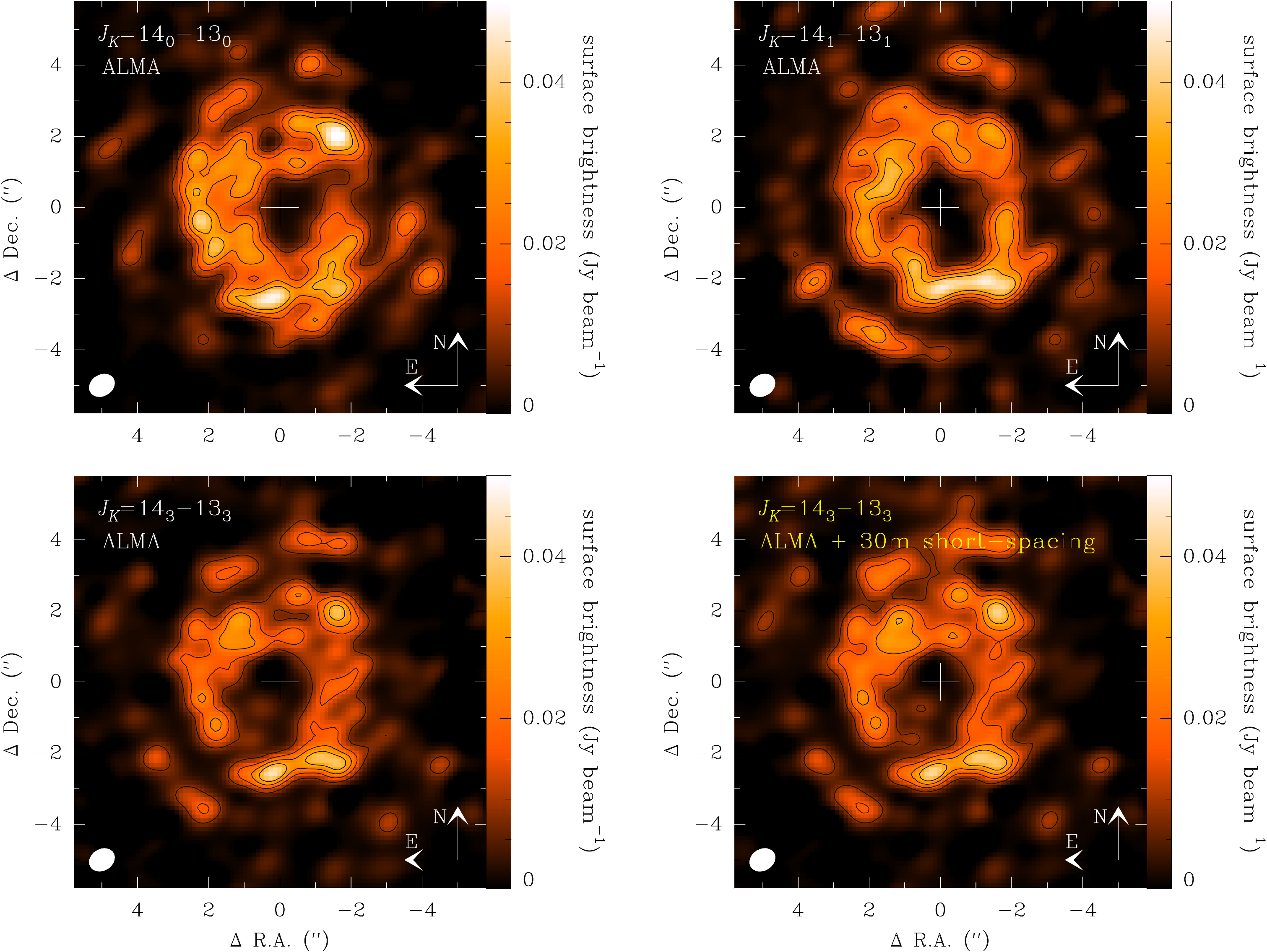}
\caption{Observed brightness distributions for the $K=0,1,3$ components of CH$_3$CN $J=14-13$, averaged over a velocity range of width 2.8~km~s$^{-1}$ centered at $V_{LSR}=V_{sys}$. Since each line has a width of $\sim$29~km~s$^{-1}$, the $K$=0 and $K$=1 components overlap and therefore the $J_K=14_0-13_0$ map contains a contribution from the transition $K=1$ line at a velocity $V=-5.8$~km~s$^{-1}$ while the $J_K=14_1-13_1$ map is also contaminated by emission from the $K=0$ component at $V=+5.8$~km~s$^{-1}$. The $J_K=14_3-13_3$ map (also shown after merging ALMA and 30m short-spacing data; bottom right panel) is free of contamination and therefore traces better the spatial distribution of CH$_3$CN in the plane of the sky. The rms per 2.8~km~s$^{-1}$ channel is 1.6~mJy~beam$^{-1}$ and contour levels are shown at 5, 10, 15, and 20 times the rms. The maps are centered on the star, indicated by a cross and with coordinates J2000.0 R.A.=09$^{\rm h}$47$^{\rm m}$57\rasecp445, Dec.=13$^{\circ}$16$'$:43\secp86. The size and shape of the synthesized beam (0\secp76 $\times$ 0\secp61) is shown in the bottom left corner of each panel.} \label{fig:map}
\end{figure*}

Some conclusions can be drawn after inspecting the brightness distribution in the different channels covering the most intense $K$ components ($K=0-3$) of the CH$_3$CN $J=14-13$ transition. Hereafter, we refer to $V_{LSR}-V_{sys}$, i.e., the local standard rest velocity corrected for the systemic velocity of the source, as simply $V$. In the case of IRC\,+10216, the systemic velocity $V_{sys}$ is $-$26.5~km~s$^{-1}$ and the terminal expansion velocity of the envelope is 14.5~km~s$^{-1}$ \citep{cer2000}. At velocities close to the terminal expansion velocity, the emission appears compact and centered on the star while at velocities around $V_{LSR}=V_{sys}$ the emission is distributed as a ring around the central position. This is the typical pattern of a molecule which is not concentrated around the central star but distributed in a hollow shell at a certain distance from the star. We must however note that in the case of CH$_3$CN $J=14-13$, this pattern is less clearly appreciated for the $K=0$ and $K=1$ components because these two lines overlap (their rest frequencies are closer than the line width of 29 km~s$^{-1}$), and therefore some of the channels have contributions from both $K$ components at different velocities.

In Fig.~\ref{fig:map} we show the brightness distribution of the strongest $K$ components ($K=0,1,3$) of the $J=14-13$ transition of CH$_3$CN at $V_{LSR}=V_{sys}$. We first focus on the maps based on ALMA data without including short-spacing (top and bottom left panels in Fig.~\ref{fig:map}). It is seen that the maximum of the emission appears as a ring with a radius of $\sim2''$ and that there is a hole within $\sim1''$ from the position of the star. The map corresponding to the $J_K=14_0-13_0$ transition contains also emission from the $K=1$ component at $V=-5.8$~km~s$^{-1}$, which makes the ring to broaden inwards. The same occurs in the map of the $J_K=14_1-13_1$ transition, which is contaminated by the $K=0$ component at $V=+5.8$~km~s$^{-1}$. In the case of the $K=2$ and $K=3$ lines, emission at $V_{LSR}=V_{sys}$ is free of blending with other $K$ components. In particular, the map of the $J_K=14_3-13_3$ transition, which is more intense than the $J_K=14_2-13_2$ transition, allows to see clearly the ring structure. It is also worth to note that a second ring of emission located at $\sim4''$ from the star is barely apparent at a few $\sigma$ in the $J_K=14_1-13_1$ map, suggesting a spiral structure which has been also observed with ALMA in other molecular lines (Cernicharo et al. in preparation) and at large scales in the $J=2-1$ line of CO \citep{cer2015}, and that has been interpreted in terms of a binary system.

To evaluate whether the brightness distributions of the different $K$ lines at $V_{LSR}=V_{sys}$ could be significantly affected by the ALMA flux loss, we have combined the ALMA data with the IRAM 30m short-spacing data to include visibilities at shorter baselines than the shortest one in the ALMA data (15.7 m). We however note that, given the modest sensitivity of the IRAM 30m data and the fact that the CH$_3$CN emission is restricted to a region much smaller than the primary beam of ALMA ($\sim23''$), we do not expect great changes in the emission structure. As an example, we compare in the bottom panels of Fig.~\ref{fig:map} the brightness distribution of the $J_K=14_3-13_3$ line at $V_{LSR}=V_{sys}$, as obtained using ALMA data alone and after combining ALMA and 30m short-spacing data. It is seen that, appart from a slight recovery of flux at extended scales, the emission structure remains essentially unaltered.

\begin{figure}
\includegraphics[angle=0,width=\columnwidth]{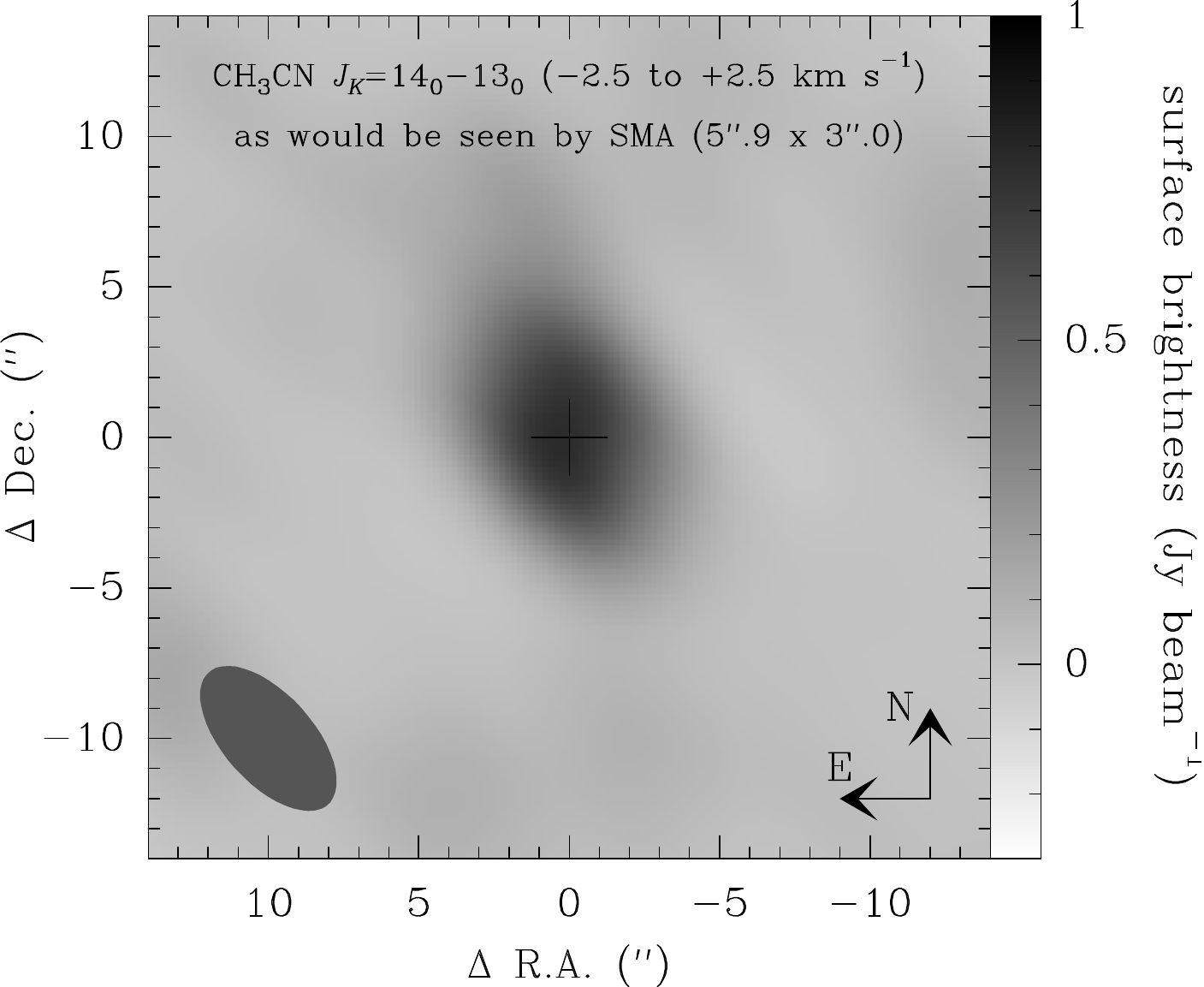}
\caption{Brightness distribution of CH$_3$CN $J_K$=14$_0$-13$_0$, averaged over a velocity range $V$ from $-$2.5 to +2.5~km s$^{-1}$, as would be seen with a synthesized beam of 5\secp9 $\times$ 3\secp0 (shown in the bottom left corner). To be compared with the brightness distribution of the CH$_3$CN $J_K$=$16_0$-$15_0$ transition observed by \citet{pat2011} with SMA (line 9 in their Figure~5). Note that with this angular resolution the inner hole of emission seen with ALMA is not resolved.} \label{fig:map-sma}
\end{figure}

Higher $J$ lines of CH$_3$CN were observed in IRC\,+10216 with the Submillimeter Array (SMA) by \citet{pat2011} in their spectral survey covering the frequency range 293.9-354.8~GHz. These authors find a compact brightness distribution for the CH$_3$CN lines, observed at angular resolutions ranging from $2''$ to $6''$, i.e., noticeably poorer than the angular resolution of the ALMA observations presented here. Moreover, these authors assign various emission lines to rotational transitions of CH$_3$CN in the $\nu_8=1$ vibrational state on the basis of a good agreement between the observed and laboratory rest frequencies. These results may lead to the conclusion that in IRC\,+10216, CH$_3$CN is present in the warm surroundings of the AGB star, as occurs with other molecules such as SiS \citep{fon2006,agu2012,vel2015}, which is in conflict with our results. On the one hand, the identification of CH$_3$CN $\nu_8=1$ by \citet{pat2011} is not convincing taking into account that for many $J$ transitions, $K$ components that should be more intense than those identified are missing. For example, in the case of the $J=16-15$ transition in the $\nu_8=1$ state, only the $K=11,12,13$ components are identified, but not the $K=0,1,2,3$ lines, which should be far more intense. Therefore, the lines assigned to CH$_3$CN $\nu_8=1$ must arise from other carriers. On the other hand, as concerns the rotational lines of CH$_3$CN in the ground vibrational state observed with SMA, the compact emission distribution observed is probably a consequence of the limited angular resolution, which does not allow to resolve the inner hole seen by ALMA. This is illustrated in Fig.~\ref{fig:map-sma}, where we have applied a tapering procedure to the ALMA data to suppress long baselines and simulate what would be seen with a synthesized beam of 5\secp9 $\times$ 3\secp0, as in the SMA observations of, e.g., the $J_K=16_0-15_0$ (see \citealt{pat2011}). It is seen that with such limited angular resolution the inner hole in the emission is not resolved. Moreover, since the CH$_3$CN lines observed by SMA are higher in $J$ than the one observed with ALMA, their emission could be shifted inwards due to excitation requirements, and thus a higher angular resolution would be needed to resolve the inner hole. We thus conclude that the hollow shell structure seen by ALMA in CH$_3$CN $J=14-13$ emission is not in conflict with the compact distribution observed for higher $J$ transitions with SMA.

\section{Discussion} \label{sec:discussion}

According to the interferometric maps carried out during the past 20 years, the molecular emission in IRC\,+10216 appears either concentrated around the central star, as in the case of HCN, CS, SiO, SiS, and NaCN, or distributed in a hollow shell located at a radius of $\sim15''$ from the star, as occurs for HC$_5$N, HNC, MgNC, and the radicals CN, C$_2$H, C$_3$H, and C$_4$H \citep{luc1995,day1995,gue1997,fon2014,vel2015}. However, in the case of CH$_3$CN, the shell is not located at $\sim15''$ from the star but at just $\sim2''$.

\subsection{Excitation and radiative transfer model}

\begin{table}
\caption{Vibrational modes of CH$_3$CN} \label{table:vibration}
\centering
\begin{tabular}{lccc}
\hline \hline
\multicolumn{1}{l}{Vibrational mode} & \multicolumn{1}{c}{Sym.$^a$} & \multicolumn{1}{c}{Freq.$^b$} & \multicolumn{1}{c}{$A$$^b$} \\
\multicolumn{1}{c}{}          & \multicolumn{1}{c}{} & \multicolumn{1}{c}{(cm$^{-1}$)} & \multicolumn{1}{c}{(s$^{-1}$)} \\
\hline
$\nu_1$ CH$_3$ symmetric stretching & A & 2954  & 3.1 \\
$\nu_2$ CN stretching & A & 2268  & 1.1 \\
$\nu_3$ CH$_3$ symmetric deformation & A & 1385.2  & -- \\
$\nu_4$ CC stretching & A & 920.290  & 0.21 \\
$\nu_5$ CH$_3$ antisymmetric stretching & E & 3009  & 1.0 \\
$\nu_6$ CH$_3$ antisymmetric deformation & E & 1449.7  & 5.3 \\
$\nu_7$ CH$_3$ rocking & E & 1041.855 & 0.50 \\
$\nu_8$ CCN bending & E & 365.024  & 0.016 \\
\hline
\end{tabular}
\tablecomments{Table built from the compilation by J. Crovisier at \texttt{http://www.lesia.obspm.fr/perso/jacques-crovisier/basemole} and from Table~1 of     \citet{mul2015}.\\
$^a$ Symmetry of first excited state of each mode ($\nu_i=1$). \\
$^b$ Frequency and Einstein coefficient for spontaneous emission of the fundamental band of each mode ($\nu_i=1\rightarrow0$).}
\end{table}

\begin{figure}
\includegraphics[angle=0,width=\columnwidth]{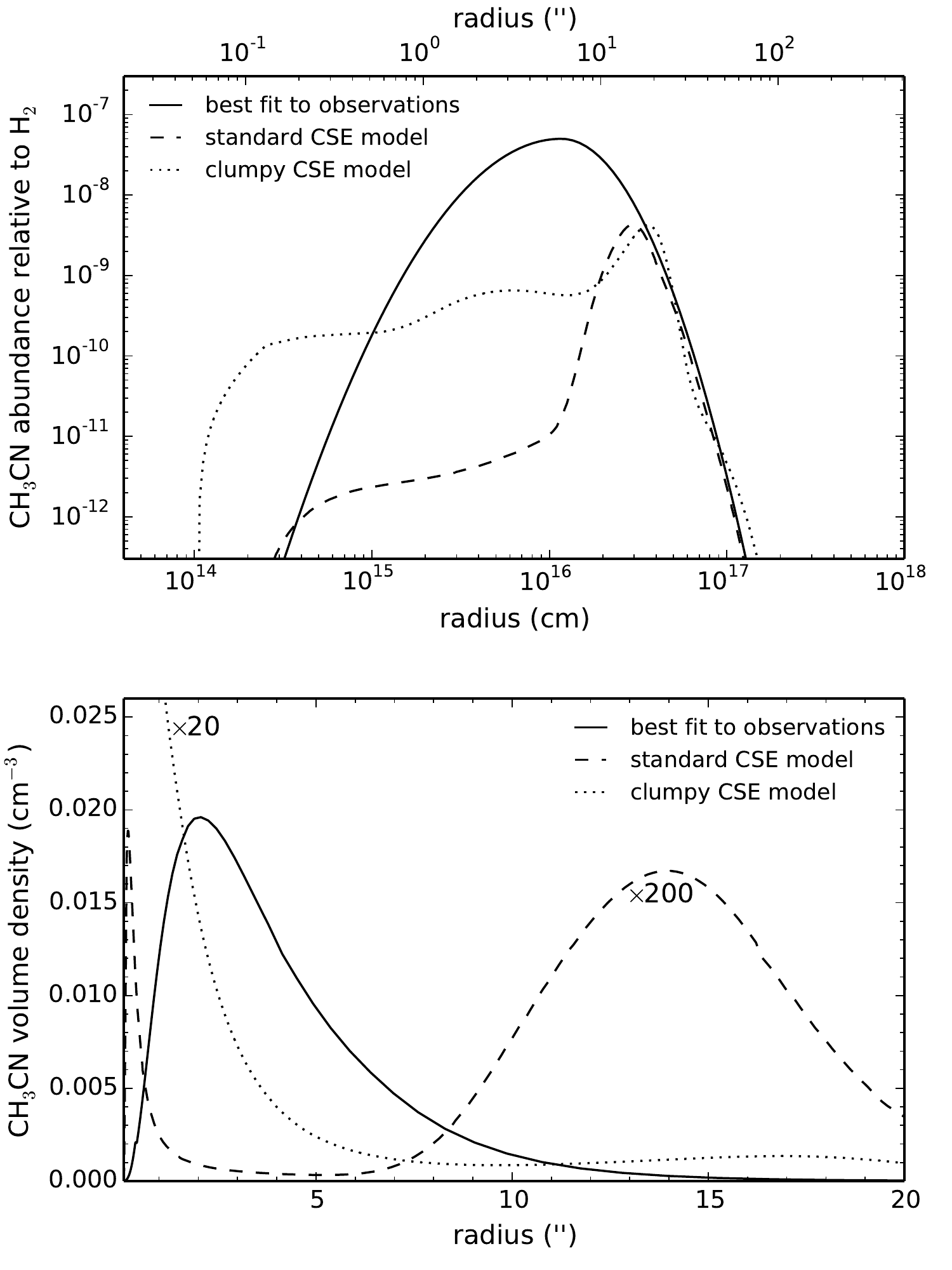}
\caption{Radial abundance profiles of CH$_3$CN in IRC\,+10216 are given as fractional abundance relative to H$_2$ in the upper panel and as volume density, scaled in some cases, in the lower panel. Solid lines represent the abundance profile that best reproduces the observations, dashed lines the result of a standard chemical model of the envelope, and dotted lines refer to a photochemical model of a clumpy envelope. The radius is given in arcseconds for an assumed distance of 130~pc.} \label{fig:abundance}
\end{figure}

In order to use the ALMA observations to put constraints on the abundance, distribution, and excitation of CH$_3$CN in the envelope of IRC\,+10216 we have carried out excitation and radiative transfer calculations based on a multi-shell LVG formalism. We adopted the model of the envelope of \citet{agu2012}, with the downward revision of the density of particles derived by \citet{cer2013} for the innermost layers ($<5R_*$). The adopted mass loss rate and distance are $2\times10^{-5}$~M$_{\odot}$~yr$^{-1}$ and 130~pc (see Table~9 of \citealt{agu2012} for other parameters). As concerns the spectroscopy of CH$_3$CN we consider separately the A and E species with the A:E statistical ratio of 1:1 and compute the energy levels from the spectroscopic constants reported by \citet{sim2004} and \citet{caz2006,caz2008} and the line strengths from the dipole moment of $3.92197\pm0.00013$~D measured by \citet{gad1995}. Infrared pumping to excited vibrational states may also play a role in the excitation of CH$_3$CN. This molecule has eight vibrational modes, whose fundamental bands lie at wavelengths from 3.3~$\mu$m to 27~$\mu$m (see Table~\ref{table:vibration}), and the radiation field in the highly reddened envelope of IRC\,+10216 is intense in the mid-infrared, especially around 10~$\mu$m \citep{cer1999}. A detailed analysis of the complete pumping scheme of CH$_3$CN is beyond the scope of this article. However, to have an idea of the effect of infrared pumping we included the first excited state of the vibrational mode $\nu_6$, which can be pumped by 6.9~$\mu$m photons. Other vibrational bands could also be important, e.g., the fundamental band of $\nu_3$, whose strength is unfortunately not available, and overtones and hot bands of low-frequency modes such as $\nu_8$. Vibrational modes such as $\nu_1$, $\nu_2$, or $\nu_5$ have also strong fundamental bands, although they lie at shorter wavelengths where the radiation field in IRC\,+10216 is less intense. The spectroscopic constants of the $\nu_6=1$ state have been taken from \citet{pas1994}. The strength of the vibrational band $\nu_6=0\rightarrow1$ has been measured by \cite{cer1985}. We considered rotational levels within the ground and $\nu_6=1$ vibrational states up to $J=30$. We adopted the rate coefficients of pure rotational excitation through inelastic collisions with H$_2$ from the calculations by \citet{gre1986}, which extend up to 140 K, and extrapolated at higher temperatures assuming that the rate coefficients scale with the square root of temperature. Collisional excitation for ro-vibrational transitions was neglected.

As radial abundance profile for CH$_3$CN we used an expression of Gaussian type which takes the form
\begin{equation}
f = f(r_0) \exp\Big\{-\Big(\frac{\log r- \log r_0}{\Delta \log r}\Big)^2\Big\}, \label{eq:abundance}
\end{equation}
where $f$ is the fractional abundance with respect to H$_2$, $r_0$ is the radius where $f$ peaks, and $\Delta \log r$ is a measure of the radial extent. The radial abundance profile which best reproduces the $J=14-13$ transition observed with ALMA is given by Eq.~(\ref{eq:abundance}) with $r_0=1.2\times10^{16}$~cm, $f(r_0)=5\times10^{-8}$, and $\Delta \log r=0.45$ for $r<r_0$ and 0.3 for $r>r_0$ (see solid lines in Fig.~\ref{fig:abundance}). It is seen that the fractional abundance relative to H$_2$ reaches a maximum value of $5\times10^{-8}$ at 6$''$ (see solid line in upper panel of Fig.~\ref{fig:abundance}). However, because of the increase in the gas density with decreasing radius, if the abundance is expressed as number of CH$_3$CN molecules per unit volume, the peak value is reached well before, at about 2$''$ (see solid line in lower panel of Fig.~\ref{fig:abundance}).

\begin{figure}
\includegraphics[angle=0,width=\columnwidth]{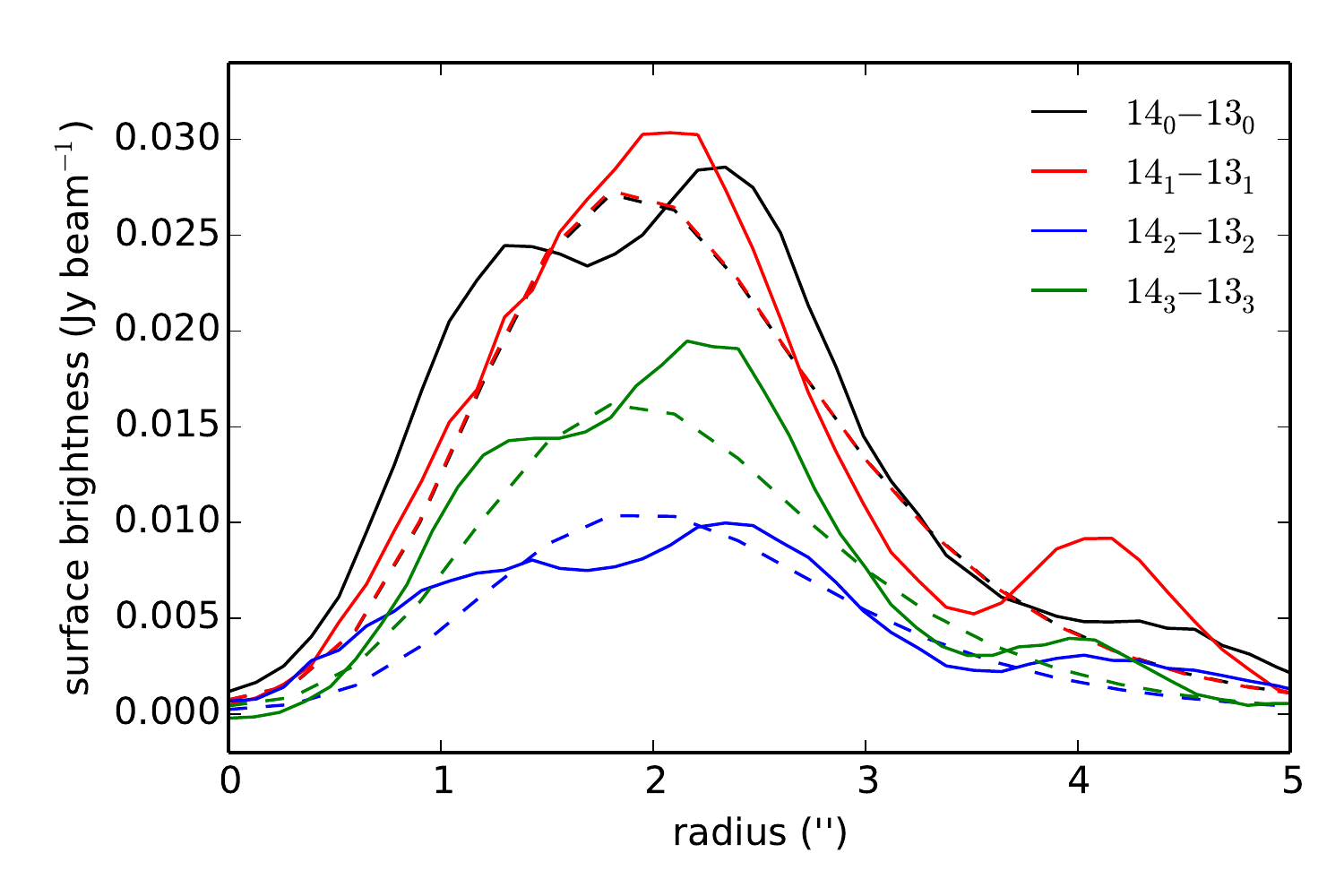}
\caption{Radial brightness distribution of the $K=0$ to $K=3$ components of CH$_3$CN $J=14-13$, averaged over a velocity range of width 2.8~km~s$^{-1}$ centered at $V_{LSR}=V_{sys}$. Note that the $K=0$ and $K=1$ components contain also a contribution from $K=1$ at $V=-5.8$~km~s$^{-1}$ and from $K=0$ at $V=+5.8$~km~s$^{-1}$, respectively. Solid lines correspond to the ALMA data combined with the IRAM 30m short-spacing data and dashed lines to the results of the model that best reproduces the observations (black and red dashed lines are almost indistinguishable).} \label{fig:azimuth_average}
\end{figure}

The radial brightness distributions calculated for the various $K$ components of the $J=14-13$ transition are confronted with the observed ones in Fig.~\ref{fig:azimuth_average}. According to our calculations, the excitation of the rotational levels involved in the $J=14-13$ transition is dominated by inelastic collisions with H$_2$, although infrared pumping to the $\nu_6=1$ state does also play a role as it induces a slight increase in the line intensity of about 10 \% and shifts the emission to slightly outer radii. Including other vibrationally excited states could enhance the importance of infrared pumping. We find that as the quantum number $K$ increases, and so does the upper level energy of the corresponding transition, the excitation requirements become more demanding and the maximum of brightness shifts to shorter radii, where the higher gas density and infrared radiation field favor a more efficient excitation. This effect is however noticeable for high$-K$ transitions. In the case of the $K=0$ to $K=3$ components, all them show a brightness peak at about 2$''$, the same radius at which the volume density of CH$_3$CN molecules reaches its maximum. It is also worth noting that the observed radial brightness distribution of three out of the four $K$ components shown in Fig.~\ref{fig:azimuth_average} show a more or less marked double-humped shape, which probably arises from the non-uniform character of the CH$_3$CN shell, with subshells at smaller scales, spiral-like structures, etc. Our model of a smooth hollow spherical shell does not aim at reproducing these fine details, but the global shape of the observed brightness distribution.

In any case, it is clear that the CH$_3$CN shell is located in IRC\,+10216 at a much shorter radius than the molecular shell at $\sim$15$''$, where the radicals CN, C$_2$H, C$_3$H, and C$_4$H, and molecules such as HNC and HC$_5$N are observed \citep{gue1997}.

\subsection{Chemical model}

Standard chemical models of IRC\,+10216 cannot reproduce the ALMA observations in that they predict that CH$_3$CN forms at too large radii \citep{agu2008,li2014}. Here we have constructed a standard chemical model using a gas phase reaction network built up from the databases UMIST and KIDA \citep{mce2013,wak2015} and the literature on gas phase chemical kinetics. We have chosen an initial radius of $2\times10^{14}$~cm, which corresponds to 0\secp1 for an assumed distance of 130 pc, to investigate the possible formation of CH$_3$CN in the inner regions of the envelope. The initial abundances injected into the expanding envelope are taken from \cite{agu2012} and N$_2$ is assumed to have an abundance of $4\times10^{-5}$ relative to H$_2$.

According to the chemical model, the formation of methyl cyanide throughout the envelope of IRC\,+10216 (see also \citealt{agu2008}) involves the formation of the precursor ion CH$_3$CNH$^+$ through the radiative association reaction
\begin{equation}
\rm CH_3^+ +  HCN \rightarrow CH_3CNH^+ + {\textit h}\nu, \label{eq:reaction-ra}
\end{equation}
followed by the dissociative recombination of CH$_3$CNH$^+$ with electrons. Among the reagents in reaction~(\ref{eq:reaction-ra}), HCN is fairly abundant from the innermost layers out to the photodissociation region \citep{fon2008,cer2011} while the ion CH$_3^+$ appears from the first stages of the expansion driven by cosmic rays although the bulk of formation is induced by photochemistry in the outer layers. The calculated abundance of CH$_3$CN relative to H$_2$ is shown as a function of radius in the upper panel of Fig.~\ref{fig:abundance} (dashed line). Note the early appearance driven by cosmic rays at radii shorter than $\sim10^{16}$~cm and the formation bump induced farther out by photochemistry. As previous chemical models, ours also puts most of CH$_3$CN in the outer envelope, in a shell located at about 15$''$ from the star. It is however interesting to note that the non-negligible fractional abundance of CH$_3$CN driven by cosmic rays from the inner layers together with the high gas densities prevailing in these regions lead to a distribution of CH$_3$CN molecules which, appart from the component located at 15$''$, has also an inner component within 1$''$ from the star. This is more clearly appreciated when expressing the abundance in absolute rather than relative terms (dashed line in lower panel of Fig.~\ref{fig:abundance}). The finding of the inner component of methyl cyanide in our chemical model is interesting because it appears closer to the region where this molecule is observed than the outer shell at 15$''$. However, this inner component would appear compact rather than showing an inner hole, for angular resolutions poorer than $\sim$0\secp1, as is the case of our ALMA data.

Our chemical model underestimates the abundance of CH$_3$CN by more than two orders of magnitude and predict a radial distribution markedly different from that observed. There are however uncertainties in the chemical model which are worth to discuss. Concerning reaction rate constants, that of the dissociative recombination of CH$_3$CNH$^+$ with electrons is known from measurements involving the fully deuterated ion \citep{vig2008}, although the branching ratios of the different channels, including that yielding CH$_3$CN, are based on simple guesses \citep{vig2009}. The rate constant of reaction~(\ref{eq:reaction-ra}) is not well constrained. Low pressure experiments suggest a value of $2\times10^{-10}$~cm$^3$~s$^{-1}$ at 300~K \citep{ani1993} while statistical calculations point to a value 45 times higher \citep{her1985}. In the chemical model we adopt the lower experimental value. If the higher theoretical value is adopted, the calculated abundance of CH$_3$CN experiences a ten-fold enhancement, although the shape of the radial profile remains similar. Another important parameter that affects the abundance of methyl cyanide, especially in the regions inner to $10^{16}$~cm, is the cosmic-ray ionization rate of H$_2$, which in the chemical model is set to a value of $1.2\times10^{-17}$~s$^{-1}$. If we adopt a higher rate, the abundance of CH$_3$CN increases at short radii. That is, the inner component gains importance with respect to the outer one. In any case, the ionization rate due to cosmic rays in IRC\,+10216 is relatively well constrained by observations of HCO$^+$ at millimeter wavelengths \citep{agu2006,pul2011}, and therefore cannot be much higher than a few $10^{-17}$~s$^{-1}$.

As long as a standard chemical model fails to explain the distribution of CH$_3$CN in IRC\,+10216, it seems clear that such model misses either some key chemical process or an adequate description of the physical structure of the envelope at arcsecond scales. There are various aspects which are worth to discuss. It is known that the envelope IRC\,+10216 is not completely smooth, but that it consists of shells and arcs which cross each other \citep{mau2000,cer2015}. The effect of these shells on the radial distribution of molecules in IRC\,+10216 has been studied by \citet{cor2009}, although the case of CH$_3$CN was not specifically discussed by these authors. A non-uniform radial density profile could modulate the early appearance of CH$_3$CN in the envelope, making the inner component to appear as a small hollow shell rather than a compact structure. To investigate this possibility it is necessary to have a precise knowledge of how the gas is distributed in the inner circumstellar layers. Another aspect of having a non-uniform distribution of matter in the circumstellar envelope is the clumpy nature, which may allow interstellar UV photons to penetrate in the inner layers, allowing to form some molecules as a result of photochemistry \citep{dec2010,agu2010}. In Fig.~\ref{fig:abundance} we show the expected distribution of CH$_3$CN under such scenario adopting the same parametric treatment of clumpiness of \cite{dec2010}. It is seen that the abundance of CH$_3$CN indeed increases in the inner layers (dotted lines in Fig.~\ref{fig:abundance}), although the distribution of molecules would appear compact around the star rather than as a hollow shell. It remains to be seen whether a different scenario of clumpy envelope or a combination of clumpiness and shells could result in an abundance distribution of CH$_3$CN compatible with that observed by ALMA. Alternative formation mechanisms such as grain surface processes could also play a role in the synthesis of CH$_3$CN. Such possibility, although difficult to prove, is an interesting one because CH$_3$CN appears soon after the dust condensation region in IRC\,+10216. Additional clues to understand the puzzling distribution of CH$_3$CN in IRC\,+10216 could come from further ALMA observations of other species showing similar distributions. In particular, it would be interesting to see whether the chemically related species CH$_2$CN shows a similar spatial distribution, although the lines of this radical are perhaps too faint to be mapped \citep{agu2008}.

\section{Conclusions}

We used ALMA to observe the $J=14-13$ rotational transition of CH$_3$CN, lying at 257.5 GHz, with an angular resolution of $0\secp76\times0\secp61$ in the carbon star envelope IRC\,+10216. The bulk of the emission is distributed as a hollow shell $\sim1''$ wide located at $\sim2''$ from the star. The observations, together with an excitation and radiative transfer analysis, indicate that methyl cyanide is not formed neither in the stellar photosphere nor far in the outer envelope, but at radial distances as short as $1-2''$, reaching a maximum abundance of $\sim0.02$~molecules~cm$^{-3}$ at 2$''$ from the star. This fact is in conflict with standard chemical models of IRC\,+10216, which predict that the bulk of CH$_3$CN molecules should be present at a radius of $\sim15''$ with an additional inner component within 1$''$ from the star. It is not yet clear which mechanism is at the origin of the peculiar spatial distribution of methyl cyanide in IRC\,+10216. The non-uniform structure of the envelope, with shells, arcs, and clumps, and grain surface processes may play an important role on the distribution of CH$_3$CN. Further ALMA observations of other molecules showing spatial distributions similar to that of CH$_3$CN should bring light on the puzzling distribution of methyl cyanide in IRC\,+10216.

\acknowledgements

We thank funding support from the European Research Council (ERC Grant 610256: NANOCOSMOS) and from Spanish MINECO through grants CSD2009-00038, AYA2009-07304, and AYA2012-32032.

\end{document}